# Composite α-μ Based DSRC Channel Model Using Large Dataset of RSSI Measurements

Hossein Nourkhiz Mahjoub, Amin Tahmasbi-Sarvestani, S M Osman Gani, and Yaser P. Fallah

***Abstract*—Channel modeling is essential for design and performance evaluation of numerous protocols in vehicular networks. In this work, we study and provide results for large-scale and small-scale modeling of communication channel in dense vehicular networks. We first propose an approach to remove the effect of fading on deterministic part of the large-scale model and verify its accuracy using a single transmitter-receiver scenario. Two-ray model is then utilized for path-loss characterization and its parameters are derived from the empirical data based on a newly proposed method. Afterward, we use $\alpha$-$\mu$ distribution to model the fading behavior of vehicular networks for the first time, and validate its precision by Kolmogorov-Smirnov (K-S) goodness-of-fit test. To this end, the significantly better performance of utilizing $\alpha$-$\mu$ distribution over the most adopted fading distribution in the vehicular channels literature, i.e. Nakagami-$m$, in terms of passing K-S test has been investigated and statistically verified in this paper. A large received signal strength indicator (RSSI) dataset from a measurement campaign is used to evaluate our claims. Moreover, the whole model is implemented in a reliable discrete event network simulator which is widely used in the academic and industrial research for network analysis, i.e. network simulator-3 (ns-3), to show the outcome of the proposed model in the presence of upper layer network protocols.**

## I. Introduction

The accuracy of the communication channel characterization profoundly depends on the precise comprehension of the communication medium nature. An accurate channel model, in turn, has a vital role on the architectural design of communication systems and optimization of their parameters. Moreover, accurate behavior of physical layer and upper layer applications heavily rely on the fidelity of the utilized model for the communication channel.

Therefore, communication channel modeling has received a great deal of attention in the literature. In particular, modeling and analysis of wireless channels is more challenging in comparison with the wireline medium due to their more complicated nature. Complexity of the wireless channel modeling problem essentially comes from diversity and ambiguity of influential factors that affect signal propagation in the wireless media. Hence, these factors should be rigorously determined with respect to the natural characteristics of the wireless environment. For instance, parameters of the channel model for an indoor environment are known to be different from an outdoor one. This distinction is due to the significantly different movement patterns of the objects and dissimilar characteristics of the surrounding obstacles for these two mediums.

Vehicular ad hoc networks (VANETs), which have been vastly targeted by the research community, are among the most emerging outdoor technologies. Various communication technologies are under investigation by several research groups around the world in order to develop appropriate VANET standards. Standard development for Physical (PHY) and Medium Access Control (MAC) layers constitute a great portion of the ongoing VANET research activities. Dedicated Short-Range Communications (DSRC) [1], cellular networks standards such as 5G-Long Term Evolution (5G-LTE) [2], and direct Device-to-Device (D2D) protocols [3] are among the most promising candidates for this purpose.

Similar to other wireless environments, channel modeling for VANETs is also challenging due to their simultaneous exposure to several complex channel imperfections, such as path-loss, shadowing, and small-scale fading [4]-[7].

Movement patterns of transmitters and receivers, their almost equal antenna heights (in contrast with cellular networks with a base-station tower and low receiver antennas), and ad hoc nature of vehicular networks make their propagation modeling fundamentally different from cellular networks. Moreover, due to their fast changing network topology and environment, the existing models of mobile ad hoc networks (MANETs) cannot be utilized for them [8], [9]. Thus, dedicated channel models are needed to describe vehicular networks behavior at physical

Hossein Nourkhiz Mahjoub and S M Osman Gani are Ph.D. students with the Department of Electrical and Computer Engineering, University of Central Florida, Orlando, FL 32826 USA (e-mails: hnmahjoub@knights.ucf.edu, smosman.gani@knights.ucf.edu).

Amin Tahmasbi-Sarvestani is a Ph.D. candidate with the Department of Electrical Engineering and Computer Science, West Virginia University, Morgantown, WV 26506 USA (e-mail: amtahmasbi@mix.wvu.edu).

Yaser P. Fallah is an associate professor with the Department of Electrical and Computer Engineering, University of Central Florida, Orlando, FL 32826 USA (e-mail: yaser.fallah@ucf.edu).

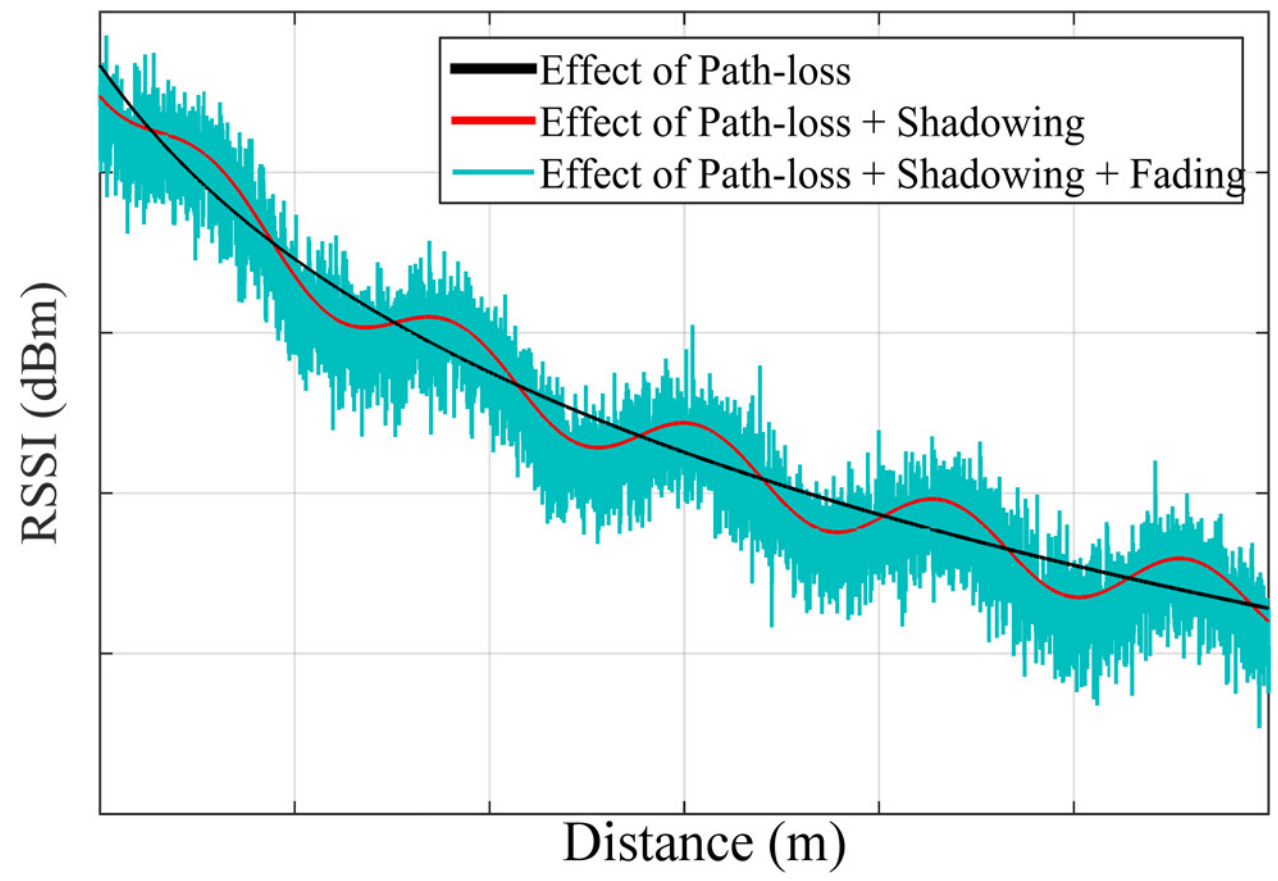

Figure 1- Effective Channel Imperfections in VANETs.

channel level.

In addition, it is more reasonable that accurate channel models be separately derived for various vehicular environments, such as highway, urban, and suburban, due to their different network characteristics. In this way, a higher degree of analysis precision will be offered when the PHY layer or upper layer protocols and applications are subject to validation through a reliable simulation environment [8], [10]-[16].

### A. Channel Attenuation Mechanisms

The main imperfections of vehicular channels in different vehicular scenarios are surveyed in [17]-[19] which can be categorized as (Figure 1)

- Path-loss, which is the large-scale propagation model, characterizes the amount of attenuation on the average signal power at the receiver. This attenuation is usually presented in terms of transmitter-receiver distance. In general, deterministic functions may be utilized to represent the path-loss effect in wireless communication channels. However, some of the works in the literature tried to propose a probabilistic path-loss model, [20]-[22], which is out of scope of this paper.
- Shadowing or large-scale fading, which is a nondeterministic model, describes the variations in the received signal power over distances which are much greater than the signal wavelength. Shadowing mainly happens in non-line-of-sight (NLOS) scenarios due to obstruction and yields to strong attenuation and variation of signal strength. The dimensions of involved obstacles are considerably greater than the wavelength of the signal, such as buildings, trees, and hills.
- Small-scale fading is a stochastic representation of diffraction, reflection, and scattering effects on the transmitted signal. The repercussion of these effects is a set of multiple waves arriving from different directions to the receiver antenna with various amplitudes and phases. It can be noted that diffraction, reflection, and scattering are consequences of transmitter or receiver movements.

Other channel characteristics such as power delay profile (PDP) and Doppler spread are also studied for the vehicular networks in the literature, which are out of scope of this paper [23].

### B. VANET Channel Modeling Data Collection Strategies

Parameters of vehicular channel models are mainly derived from utilizing two main measurement methods: channel sounder or received signal strength indicator (RSSI). On one hand, channel sounders are specifically designed to determine the frequency response of the channel. However, they are mainly utilized in small scale data measurement campaigns, since the channel characterization of situation of numerous transceivers is not feasible with channel sounders [24]-[29]. This is due to the very high cost and complexity of such experiments. On the other hand, RSSI values can be extracted merely from the received packets with no extra cost or setup. It is worth mentioning that the data which comes from the small scale measurement campaigns may not be an appropriate representative of the actual channels due to the scalability issues. For instance, since each vehicle is not only a transceiver, but also an obstacle, increasing the scenario size may drastically change the physical environment characteristics. Therefore, RSSI-based methods may become more suitable for parameter identification of actual channels with high number of vehicles, even though they are not primarily recorded to be used for channel modeling.

### C. Paper Contributions

- In this paper, using a large dataset of RSSI values collected from a realistic vehicular measurement campaign, in the first step a new approach is proposed to derive path-loss parameters for the *two-ray* propagation model. This method improves the accuracy of the achieved parameter values by taking into account the mean (expected value) of fading. Considering the fact that the mean of fading has a noticeable effect on the average received signal power, the calculated values for path-loss parameters are not precise enough without this modification on the raw RSSI values. In fact, the path-loss model parameters would indirectly be affected and biased by the fading and implicitly convey its effect if the proposed modification approach be ignored.
- Subsequently, $\alpha$-$\mu$ distribution is adopted in the context of VANETs for the first time to model the small-scale fading of the vehicular environment by the virtue of this powerful distribution.
- Afterwards, the goodness-of-fit of the proposed distribution, i.e. $\alpha$-$\mu$ distribution, is statistically validated with the Kolmogorov-Smirnov (K-S) test. We statistically demonstrated the substantial better performance of $\alpha$-$\mu$ distribution as the vehicular channel fading model, compared to the Nakagami-$m$ model. It is worth mentioning that Nakagami-$m$ model has been widely accepted as the best model and state-of-the-art for vehicular channel fading in the literature so far. To the best of our knowledge, neither modeling of fading with $\alpha$-$\mu$ distribution nor statistical validation of the goodness-of-fit for any fading model have ever been done in dense vehicular environments.

- Finally, we take our model evaluation one step further by testing its results under presence of upper layer protocols in a credible network simulator, namely network simulator-3 (ns-3), which is a realistic simulation environment for network analysis.

The rest of this paper is organized as follows. In Section II, related works are surveyed. Section III is devoted to the explanation of measurement campaign configuration. A detailed description of our propagation modeling method is presented in Section IV. The model evaluation and ns-3 simulation results are presented in Section V. Finally, we conclude the paper with a summary in SectionVI.

## II. Related Works

Different measurement campaigns have been established and numerous scenarios were investigated with respect to various surrounding environments of VANETs to identify the aforementioned parameters of vehicular channels based on both channel sounder devices and RSSI-based measurements.

Two different path-loss models are proposed for inter-vehicle communication channel for three different environments: rural, highway, and urban [24]. Measurement framework was set up at 5.9 GHz, using a channel sounder at the transmitter side which generated a multi-tone signal and one receiver. A two-ray model with a zero mean log-normal shadowing is derived as a rough path-loss model for both rural and highway environments for distances larger than 32 meters. For urban scenario, a general log-distance model with a zero mean log-normal shadowing is recommended. The path-loss exponent of 1.61 is calculated based on their measured dataset.

Path-loss, power delay profile and delay Doppler spectrum analysis in highway, rural, urban and suburban environments are investigated using channel sounder [25]-[28]. Measurements were performed over the 240 MHz band around 5.2 GHz central frequency. Two GPS-equipped trucks were employed as transmitter and receiver. The most important characteristic of all of their measurement scenarios is the dominant line-of-sight (LOS) between transmitter and receiver. Log-distance model plus a zero mean log-normal shadowing is proposed for highway, urban, and suburban environments. The parameters are evaluated for the measured data using least mean square error curve fitting method. Due to low density of scatterers, which leads to a dominant LOS plus a ground reflection, two-ray is considered to be the correct path-loss model in the rural environment. Validity of the models is claimed for distances greater than 10 and 20 meters for log-distance and two-ray, respectively.

Nakagami-$m$ and Weibull distributions were proposed in [28], [29] to model the small scale fading in a highway scenario based on the MIMO car-to-car measurements at 5.3 GHz.

In [30] two different obstruction modes are proposed as NLOS and obstructed LOS (OLOS) based on the type of obstacle between transmitter and receiver. Three different log-distance path-loss models are then derived for LOS, OLOS, and NLOS cases from a measurement data using LUND channel sounders.

Rayleigh and Rician fading models have been proposed for V2V channels with realistic non-isotropic scatterers in [31] and [32]. A novel method with higher precision for Rayleigh parameter computation was investigated in [31] while Rician envelope level crossing rate and average fade duration are investigated in [32].

A Rician geometry-based stochastic fading model has been proposed for wideband multiple-input multiple-output (MIMO) V2V channels in [33] and a method for inter-carrier interference (ICI) cancellation is proposed based on this model.

In [34], fading statistics are analyzed in four environments: highway, motorway, urban and suburban. Two vehicles, one transmitter equipped with a 5.2 GHz signal generator and one receiver equipped with a network analyzer were used for data collection. It is claimed that urban and suburban environments have the same behavior, in terms of small-scale fading, as motorway and highway scenarios, respectively. Rician, Rayleigh, Gaussian and Nakagami-$m$ distributions were examined and Rician channel model was found to be the best match in all scenarios based on the least mean absolute error method which was reasoned by dominant LOS in all measurements. Log-normal model was offered to be used for shadowing in all scenarios. It should be mentioned that flat fading assumption for vehicular channels is insupportable for DSRC-based networks [35], [36]. Therefore, due to the frequency selective nature of DSRC channel fading, whereas RSSI values represent the effect of total channel attenuation through energy measurement, their use is best limited to path-loss, shadowing, and fading analysis [36], [37].

Authors in [38]-[41] considered the intermediate vehicles between transmitter and receiver as dynamic obstacles and added them to static obstacles, like buildings, in the path-loss model calculations. The results indicate that path-loss of the vehicular channel, packet delivery ratio (PDR), latency, and jitter completely depend on the presence of LOS. Two vehicles, one transmitter and one receiver, were equipped with NEC LinkBird-MX which is a DSRC platform working based on IEEE 802.11p on 5.9 GHz frequency. RSSI and PDR values were logged in different measurement scenarios. The results supported the two-ray as the best fitted path-loss model in the presence of LOS for highway, suburban, and urban areas.

Empirical RSSI values from a measurement campaign in freeway environment, divided in ten-meter bins, were used to find the best fitted fading distribution [37]. Nakagami-$m$ distribution is proposed as an appropriate fit based on the cumulative distribution function (CDF) matching approach. Reported values of shape parameter $m$ for Nakagami-$m$ distribution lies between 1 to 1.8 for distances less than 100 meters, and 0.7 to 1 beyond it. Therefore, behavior of Nakagami-$m$ distribution is dominantly similar to Rician and Rayleigh for distances below and above 100 meters, respectively.

In another effort [42], path-loss and fading were analyzed for suburban environment based on two RSSI datasets collected from two vehicles as a transmitter-receiver pair. A dual slope log-distance path-loss model plus a zero mean log-normal shadowing is suggested to model large-scale power attenuation. Different path-loss exponents and shadowing standard deviations for distances under and above 100 meters are calculated using regression methods. Small-scale fading is modeled using Nakagami-$m$ distribution. The first dataset tends toward Rician and Rayleigh distributions for distances under and above 70 meters, respectively. The other dataset has the same behavior pattern except that the change point is around 90 meters.

A similar approach is taken to derive the log-distance path-loss and zero mean log-normal shadowing parameters from RSSI measurements in highway and rural environments [42], [43]. Due to the fewer number of obstacles, critical distance for slope change in path-loss model is reported much closer to the first Fresnel zone. Moreover, shadowing standard deviations of rural and highway environments are also less than shadowing standard deviation of suburban environment of [41], [42].

RSSI-based measurements with a single transmitter-receiver pair of vehicles in motorway, rural and urban environments have been utilized to identify the best fading model for vehicular scenarios in [43], [44]. Dual-slope log-distance path-loss with a zero mean log-normal shadowing and Nakagami-$m$ fading models are derived as the best match for large- and small-scale received power variations, respectively.

Nakagami-$m$ distribution is also selected to model the small-scale fading of the vehicle-to-vehicle (V2V) channel based on the RSSI measurements in a highway and an open space area [44], [45]. The results propose greater values of Nakagami-$m$ shape parameter $m$ for open space in comparison with highway.

A simulation based study of IEEE 802.11p also takes Nakagami-$m$ fading to model the vehicular networks channel [46]. A suburban scenario is simulated in OPNET with an updated physical layer model which has Nakagami-$m$ fading model included. They found the best value for shape parameter $m$, as a function of separation distance and relative velocity of transmitter and receiver.

In our work, utilizing a large dataset of RSSI values, the following approach has been taken to find appropriate models for both path-loss and small-scale fading in vehicular environments, especially in highway scenarios. In the first step, two-ray propagation model is selected to model the path-loss and its parameters are derived from our dataset. We have considered the effect of fading mean on path-loss parameters estimation process in this step. Afterwards, we use $\alpha$-$\mu$ distribution, a more general fading model compared to Nakagami-$m$, to model the small-scale fading of the vehicular environment, while its notable higher performance compared to Nakagami-$m$, which is currently the state-of-the-art fading model in the vehicular literature, is statistically validated through proper goodness-of-fit tests.

## III. Measurement Campaign

We use the dataset collected by the Crash Avoidance Metrics Partnership (CAMP) Vehicle Safety Communications 3 (VSC-3) Consortium, in partnership with the United States Department of Transportation (USDOT). This dataset was collected as part of the V2V safety communications scalability activity in a project to study the Interoperability Issues of Vehicle-to-Vehicle Based Safety Systems (V2V-Interoperability project). Although channel modeling was not the main purpose of the project, RSSI values were collected as metadata. The Atheros DSRC radios based on IEEE 802.11p standard on 5.9GHz frequency with 10MHz band were used for the measurements [47]. All of the DSRC radios logged RSSI values with 1 dB accuracy.

The field environment is a straight, flat 1400 meters 6-lane highway with 400 equidistant separated radios. Each radio transmits a fixed-length packet every 100ms. A constant transmission power is preset in all transceiver devices. They have also equal transmitting and receiving antenna gains. The RSSI values, which are recorded for a group of transceivers, are employed to model the communication channel in this work. We use two datasets of field scenarios to validate our proposed model.

It is noteworthy that conducting real-world large-scale data collections for very dense scenarios in the tackled research domain, i.e. DSRC VANETs analysis, is very difficult as it would be very expensive to put together such a huge test bed. However, based on the provided theoretical analysis of the problem and its justification throughout the presented results, it would be reasonable for the proposed model to be considered as a valid scalable solution for more challenging situations compared to the scenarios which have been analyzed in this work in terms of scenario density, different ranges of vehicle speeds, etc.

The first scenario considered in this work is the touchstone scenario by which we derive the environment-dependent parameters of two-ray path-loss model. It has also been employed to verify our proposed method for path-loss exponent evaluation (Figure 2).

The touchstone is a single transmitter-receiver scenario in which the transmitter is mounted on a vehicle at the height of 1.61 meters, while the receiver is placed on a stand with the same height at the beginning of the road where the vehicle starts. The transmitting vehicle drives away from the stationary receiver on a straight path with the constant speed of 22 km/h and turns around toward it at the distance of 1400 meters. The packet transmission rate and the transmission power are set to 10Hz and 18 dBm, respectively.

The second scenario is the *400-car* scenario in which four moving vehicles follow each other and keep 75 meters distance in the middle lane. Each vehicle is equipped with two transceivers on the roof top at the height of 1.61 meters. The rest of transceivers are placed stationary along the road at the same height on top of 65 carts. On each cart 6 transceivers are mounted to play the role of actual vehicles as shown in Figure 3. The carts are equidistant separated with 75 meters distance on four side lanes. The packet transmission rate and power of 400-car scenario are 10Hz and 20 dBm, respectively. Multiple trials of the 400-car scenario with different configurations exist which are used to verify the proposed path-loss and fading models.

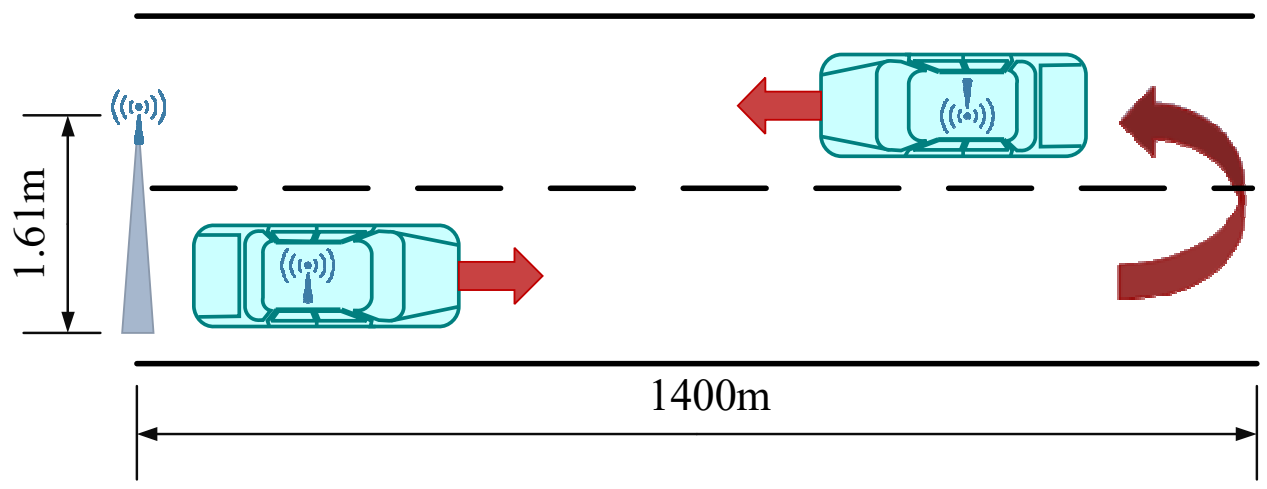

Figure 2- Touchstone Scenario

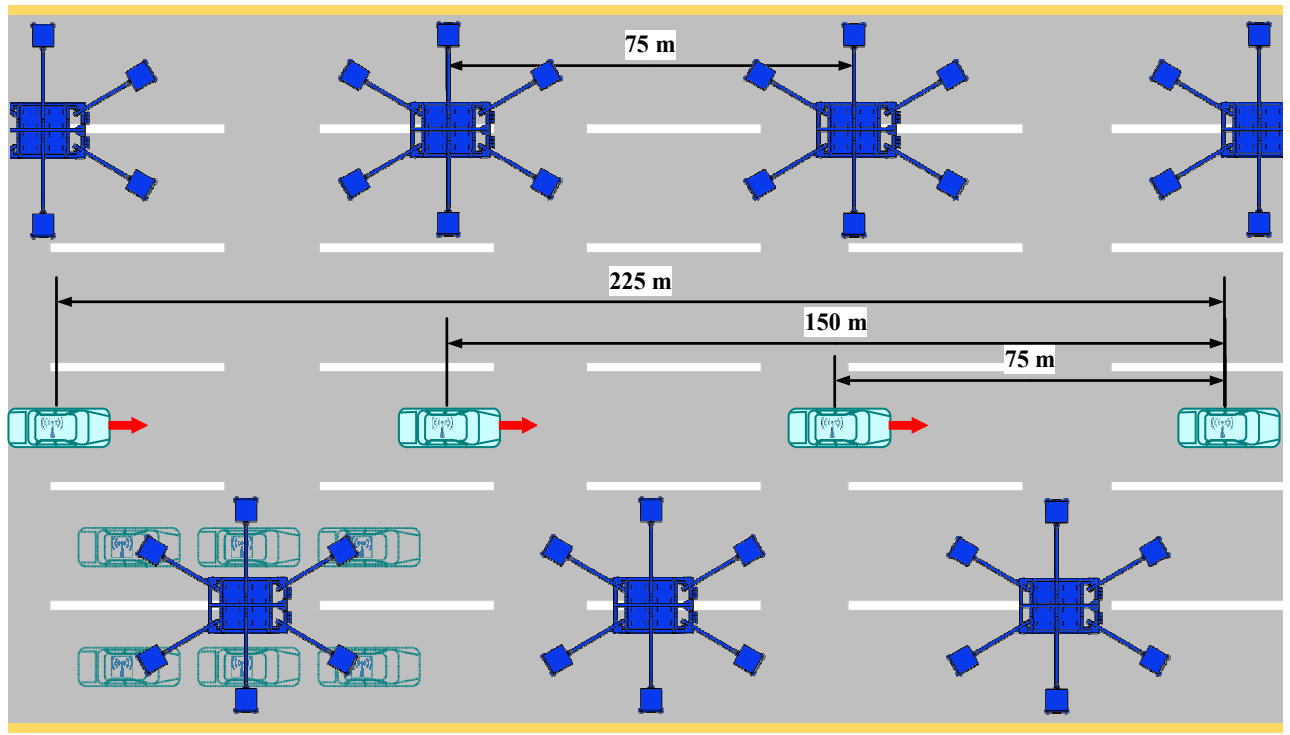

Figure 3- 400-car Scenario

## IV. Propagation Modeling Methodology

Communication channel determines how the transmitted signal is changing as it propagates. Although it is impossible to exactly identify the behavior of the channel, propagation models have tried to predict the received signal strength to a great extent. As known from the propagation model of a signal in a wireless communication channel, the received power from a transmitter is affected by three well-known factors, namely path-loss, shadowing, and fading. The logarithmic representation of received signal strength at an arbitrary distance $d$ can be written as:

$$P_r(d) = P_t - L_{LS}(d) + g_{P,dB}(dB) + X_\sigma \tag{1}$$

where $P_t$ is the transmission power in dB, $L_{LS}(d)$ is the deterministic path-loss at distance $d$, $g_{P,dB}$ is the small-scale fading, and $X_\sigma$ is the zero-mean shadowing with standard deviation of $\sigma$. Due to the fact that transceiver carts used in the data collection campaign do not represent the actual obstruction and dimension of 6 vehicles, the environment is sparse and free from any large objects to cause shadowing, except for the four vehicles. Therefore, an accurate shadowing model cannot be derived from our experiments for real highway scenarios.

Considering the setup of touchstone and 400-car scenarios with straight, flat road and vehicles with antennas of the same height, there always exist a dominant direct line-of-sight (LOS) ray and a ground-reflected ray between each transmitter-receiver pair. Therefore, it is reasonable to employ two-ray ground reflection model under these circumstances. As mentioned before, two-ray is among the best candidates to model path-loss in vehicular networks [24], [48]. The following section briefly describes the two-ray model and our proposed approach to find its parameters.

### A. Path-loss Model Derivation

As previously mentioned, two-ray ground reflection model is widely used to model path-loss in vehicular networks. This model can be formulated as [48]

$$L_{LS}(d;\gamma,\epsilon_r) = 10 \times \gamma \times \log_{10}\left(4\pi\frac{d}{\lambda}\left|1 + \Gamma_\perp e^{i\varphi}\right|^{-1}\right) \tag{2}$$

where $\gamma$ is the path-loss exponent. This parameter is introduced as an empirical environment-dependent adaptation of non-ideal channel conditions and assumed to be 2 in ideal channel conditions [48], [49]. $\lambda = \frac{c}{f}$ is the signal wavelength corresponding to the transmitted signal with central frequency of $f$ that is propagating in the environment with the speed of $c$. In the above equation, the reflection coefficient $\Gamma_\perp$ can be found as

$$\Gamma_\perp = \frac{\sin\theta - \sqrt{\epsilon_r - \cos^2\theta}}{\sin\theta + \sqrt{\epsilon_r - \cos^2\theta}} \tag{3}$$

where $\epsilon_r$ is a fixed, unit-less constant dependent on the reflection medium. $\sin\theta = \frac{h_t + h_r}{d_{ref}}$ and $\cos\theta = \frac{d}{d_{ref}}$, $d_{ref} = \sqrt{d^2 + (h_t + h_r)^2}$ are dependent on $h_t$ and $h_r$ which are the heights of the transmitter and receiver antennas, respectively. Furthermore, the phase difference of the two interfering rays, $\varphi$, can be found as

$$\varphi = 2\pi \times \frac{d_{los} - d_{ref}}{\lambda}, \tag{4}$$

where $d_{los} = \sqrt{d^2 + (h_t - h_r)^2}$ .

Therefore, the path-loss model defined in Equation (2) has two unknown environment-dependent parameters of $\gamma$ and $\epsilon_r$ that should be found based on the empirical data.

Regression methods have been employed in some previous works to determine the path-loss parameters [42], [48]. These methods try to fit a line or a non-linear curve on the empirical data points by satisfying some criteria, such as minimizing mean of the square errors (MSE). However, the envelope and fluctuation span of collected data points have a significant impact on the applicability of regression methods for path-loss model characterization. For instance, in scenarios like our touchstone scenario in which there are only few data points representing the span of small scale fading, fading mean has a negligible effect on the empirical data average. Therefore, regression methods result in an acceptable set of values for path-loss parameters, even though they do not take into account any prior knowledge about the nature of the problem and only want to satisfy their criteria over the whole batch of data points. However, for scenarios such as 400-car, in which the data points are conveying the noticeable and meaningful fluctuations due to the small scale fading, applying these methods on the raw data points without any pre-processing results in improper values for path-loss parameters. The envelope of the small scale fading in these types of scenarios substantially affect the average received signal strength values. In general, this effect varies over different transmitter-receiver separation distances. This variation is a consequence of the changes in fading distribution parameters. However, the regression methods merely try to minimize the MSE to find the best fitted curve on these averaged values and neglect such effects. This negligence potentially results in a biased set of regression-based estimated path-loss parameters which do not represent the accurate path-loss model. Hence, employing regression methods to estimate path-loss parameters in these types of scenarios have no analytical justification and some degree of modification should be applied on the raw data first. In order to address this issue, we propose the following approach to determine more accurate values of $\gamma$ and $\epsilon_r$ and assess its validity using our touchstone scenario.

The empirical average RSSI (in dBm), denoted as $\langle P_r(d)\rangle$, for each distance $d$ (in meters) is assumed to be equal to expected value of received signal strength:

$$\langle P_r(d)\rangle = E[P_r(d)] = P_t - L_{LS}(d) + E\big[g_{P,dB}(dB)\big] \tag{5}$$

As fading characteristics do not change in close distances, for each distance $d'$, which is relatively close to $d$, we have

$$\langle P_r(d')\rangle - \langle P_r(d)\rangle = L_{LS}(d) - L_{LS}(d') = 10 \times \gamma \times \log_{10}\left(\frac{d}{d'} \times \left|\frac{1+\Gamma_{\perp,d'}e^{i\varphi_{d'}}}{1+\Gamma_{\perp,d}e^{i\varphi_{d}}}\right|\right) \tag{6}$$

It is noteworthy that what is meant here by fading characteristics is the set of parameters of the fading distribution, and not the fading values themselves. These parameters could be assumed fixed in close distances. Therefore, the expected value of fading, $E\big[g_{P,dB}(dB)\big]$, which is solely function of these parameters, would be fixed over those distances, as well.

### 1) Determining $\epsilon_r$

The value of $\epsilon_r$ is determined prior to $\gamma$ as it only depends on the natural properties of reflecting surface. Thus, the touchstone scenario is used to this end as it has exactly the same environmental properties, but less complexity in comparison with the 400-car scenario. Similar to other single transmitter-receiver scenarios with no obstacles, channel condition in touchstone scenario is assumed to be ideal. Therefore, the exponent value of two-ray path-loss model is $\gamma = 2$. This assumption is based on the fact that the environment of this scenario is an open area without any obstacles and a dominant LOS component always exists. Although using Equation (6) for any arbitrary pair of distinct distances $d$ and $d'$ should theoretically yield to an identical value of $\epsilon_r$, the obtained value may vary for each pair of distances due to the empirical data imperfection. Therefore, statistical mode of the distribution of obtained $\epsilon_r$ values is chosen as the best approximation of the actual value.

### 2) Determining $\gamma$

A similar method is used to determine the value of $\gamma$ which can be derived using Equation (6) based on the figured value of $\epsilon_r$ from the preceding subsection. The touchstone and 400-car scenarios are conducted in the same field, in which the value of $\epsilon_r$ is fixed. Hence, the corresponding $\gamma$ value for each scenario can be determined using previously calculated $\epsilon_r$. Similar to $\epsilon_r$ calculation approach, the obtained value may vary for any arbitrary pair of distinct distances $d$ and $d'$ due to imperfection of empirical data. Therefore, statistical mode of the distribution of obtained values of $\gamma$ is selected as the most accurate approximation of the actual value. The derived values for both scenarios are reported in the results section. Utilizing regression method in touchstone scenario and achieving the same value of $\gamma$ and $\epsilon_r$ approves the validity of our approach. Moreover, our approach is valid in more complicated scenarios such as 400-car, whereas regression method is not reliable as it neglects the effect of fading mean ($E\big[g_{P,dB}(dB)\big]$).

## B. Small-scale Fading Analysis

The generalized Gamma or Stacy distribution is originally derived in [50]. The probability density function (PDF) of this distribution covers some of the most widely used distributions such as Rayleigh, Gamma, one-sided Gaussian, Nakagami-$m$, exponential, and Weibull as its special cases. The $\alpha$-$\mu$ distribution, a rewritten form of the generalized Gamma distribution with change of variables, was proposed by Yacoub in the wireless communication context [51], [52]. It was originally derived based on dedicated characteristics of the wireless medium, such as field non-linearity, and then recognized as an equivalent form of Stacy distribution.

The derivation approach of the $\alpha$-$\mu$ distribution, based on the wireless channel characteristics, should be elaborated in order to make it more perceptible as a wireless channel fading model. In order to derive the $\alpha$-$\mu$ fading distribution, received signal at each point is assumed to be composed of different clusters of multipath components [51], [52]. Moreover, a non-homogenous propagation environment is considered for the problem formulation. Scattered waves which belong to each single cluster are assumed to be similar in terms of their delays, while their phase values are supposed to be random. The power is also assumed identical for scattered wave elements within each cluster. The aforementioned environment non-homogeneity causes some non-linearity in the total received signal envelope. More specifically, if the received signal be comprised of $n$ multipath components, the mentioned non-linearity could be modeled so that the total received signal envelop, denoted by $X$, be the $\alpha$-root of sum of Euclidian length squares of multipath components. This non-linearity form could be written as:

$$X = \sqrt[\alpha]{\left[\sum_{c=1}^{n}|X_c|^2\right]} \tag{7}$$

$$X_c = I_c + Q_c \tag{8}$$

Where $I_c$ and $Q_c$ are the in-phase and quadrature parts of each multipath component, $X_c$, respectively. These in-phase and quadrature parts assumed to be zero mean independent Gaussian processes with equal variances of $\sigma^2$ (for all multipath components). PDF of $P = X^\alpha$ could then be easily evaluated under these assumptions as follows:

$$f_P(p; n, \sigma) = \frac{p^{n-1}}{(2\sigma^2)^n \Gamma(n)} \exp\left(-\frac{p}{2\sigma^2}\right) \tag{9}$$

Now, applying the appropriate variable change, PDF of $X$ could be derived from (9) as:

$$f_X(x; \alpha, n, \Omega) = \frac{\alpha n^n x^{\alpha n-1}}{\Omega^{\alpha n}\Gamma(n)} \exp\left(-\frac{n x^\alpha}{\Omega^\alpha}\right) \tag{10}$$

Where $\Omega$ is $\alpha$-root mean of $P$ $\left(\Omega = \sqrt[\alpha]{E[P]}\right)$.

Parameter $n$ was a positive integer number so far, indicating the number of multipath components. However, its range could be extended to positive real values since there might be nonzero correlation between in-phase and quadrature parts of multipath components or those in-phase and quadrature parts might follow a non-Gaussian distribution. Correlation among different multipath clusters could also be another reason for this parameter range extension. Based on the above discussion, the PDF of $\alpha$-$\mu$ distribution could be calculated as

$$f_X(x; \alpha, \mu, \Omega) = \frac{\alpha \mu^\mu x^{\alpha\mu-1}}{\Omega^{\alpha\mu}\Gamma(\mu)} \exp\left(-\frac{\mu x^\alpha}{\Omega^\alpha}\right), x \geq 0, \alpha > 0, \mu > 0, \tag{11}$$

Where $\mu$ is the real extension of $n$:

$$\mu = \frac{E^2[x^\alpha]}{E[x^{2\alpha}] - E^2[x^\alpha]}, \tag{12}$$

and

$$\Omega = \sqrt[\alpha]{E[x^\alpha]}. \tag{13}$$

The $\alpha$-$\mu$ distribution is one of the most general distributions which can model both fast fading and shadowing effects concurrently [53]. The derivation paradigm for this distribution, as explained earlier, is mainly proposed to overcome the wireless environment non-linearity effect due to spatial correlation of scatterer surfaces [51], [52]. The main obstacles and scatterers in vehicular networks, especially in highway scenarios [43], are other vehicles, which are noticeably correlated in terms of their spatial distribution. Therefore, environment non-linearity, arisen from this spatial correlation of scattering surfaces, should be considered in VANETs channel modeling. In this work, we considered this characteristics of the vehicular network environments. This justification leads us to employ $\alpha$-$\mu$ distribution to model the small scale fading for VANETs. Moreover, the results of our statistical tests validate the correctness of this assumption for the collected dataset. Although it seems reasonable to leverage the non-linearity property in vehicular environments, to the best of our knowledge the $\alpha$-$\mu$ distribution has never been touched in modeling vehicular channels beforehand. As mentioned before, Nakagami-$m$ distribution is now the most adopted vehicular fading model in the literature [37], [42], [44]-[46], [53], [54].

Parameter estimation of $\alpha$-$\mu$ distribution has been studied extensively in the literature [55]-[57]. The two vastly adopted approaches are maximum likelihood and moment generating function estimators [58], [59]. As the RSSI values in our scenarios are recorded in dBm and quantized to the nearest integer value, a parameter estimator of logarithmic distribution is more desirable to eliminate the numerical truncation error due to the logarithmic to linear conversion. A set of parameter estimators for logarithmic $\alpha$-$\mu$ distribution based on log-moments and least square error are proposed and mathematically analyzed in [53], [56]. These estimators have been designed to estimate parameters of $\alpha$-$\mu$ distribution (i.e. $\alpha$, $\mu$, and $\Omega$ ) leveraging another estimator, formula (17), which has been originally proposed by Stacy in [57]. We are employing these estimators, i.e. formulas (18), (19), and (20), to obtain distribution parameters of fading model for our empirical data.

The logarithmic form of the $\alpha$-$\mu$ distribution which is derived from (11) with a change of variables could be formulated as:

$$f_L(l) = \frac{1}{A} \frac{\alpha\mu^\mu}{\Omega^{\alpha\mu}\Gamma(\mu)} \exp\left(\frac{\alpha\mu}{A} l - \frac{\mu}{\Omega^\alpha} e^{\frac{\alpha l}{A}}\right), \ -\infty < l < \infty \tag{14}$$

Where $L = A\ln(X)$ $(A = \frac{20}{\ln 10}$ and $X$ is the linear $\alpha$-$\mu$ distributed random variable in (11)).

It can be noted that the results of the logarithmic representation of moment generating functions are more accurate compared to regular moments [53].

The second and third central moments of the logarithmic $\alpha$-$\mu$ distribution are used to derive the estimators:

$$\mathcal{M}_2 = E\left[\left(l - \bar{l}\right)^2\right] = \left(\frac{A}{\alpha}\right)^2 \psi'(\mu), \tag{15}$$

$$\mathcal{M}_3 = E\left[\left(l - \bar{l}\right)^3\right] = \left(\frac{A}{\alpha}\right)^3 \psi''(\mu), \tag{16}$$

where $\psi(x) = \frac{\partial \ln \Gamma(x)}{\partial x}$, $\psi'(x)$, and $\psi''(x)$ are the psi function and its first and second order derivatives, respectively [60].

The $\hat{\tau}$ estimator is defined as [56],[57]

$$\hat{\tau} = \frac{\widehat{\mathcal{M}}_2^{\frac{3}{2}}}{\widehat{\mathcal{M}}_3} \triangleq \frac{\left(\psi'(\mu)\right)^{\frac{3}{2}}}{\psi''(\mu)}, \tag{17}$$

where $\widehat{\mathcal{M}}_2$ and $\widehat{\mathcal{M}}_3$ are empirical second and third central moments of the logarithmic $\alpha$-$\mu$ distribution, respectively. The third term of (17) could be easily derived by substituting (15) and (16) into its second term. The value of $\mu$ can then be estimated as a piecewise-defined function on the different ranges of $\hat{\tau}$ as follows:

$$\hat{\mu} = \begin{cases} \hat{\tau}^2 + \frac{1}{2}, & \hat{\tau} \leq -2.85 \\ -0.0773\hat{\tau}^4 - 0.6046\hat{\tau}^3 - 0.7949\hat{\tau}^2 & \\ -2.4675\hat{\tau} - 0.9208, & -2.85 < \hat{\tau} \leq -0.6 \\ -132.8995\hat{\tau}^3 - 232.0659\hat{\tau}^2 & \\ -137.6303\hat{\tau} - 27.3616, & -0.6 < \hat{\tau} < -0.5 \end{cases} \tag{18}$$

Accordingly, $\alpha$ and $\Omega$, can be estimated as

$$\hat{\alpha} = A\sqrt{\frac{\psi'(\hat{\mu})}{\widehat{\mathcal{M}}_2}}, \tag{19}$$

$$\widehat{\Omega} = \sqrt[\hat{\alpha}]{E\left[\exp\left(\frac{\hat{\alpha}\hat{l}}{A}\right)\right]}. \tag{20}$$

## V. Model Evaluation

Calculated parameter values for suggested path-loss and fading models based on the empirical data and their statistical evaluation are presented in this section. Moreover, the ns-3 simulation results are shown to evaluate the performance of the model with presence of upper layer protocols.

It can be noted that the DSRC radio carrier-sensitivity has been tested by the radio vendor, and it turns out that the received packets with RSSI values less than -94 dBm cannot be correctly detected and most of them are lost. Therefore, there are very few number of available data points for higher distances in the empirical observations, and the data itself is not representative since it has been pruned by the DSRC receivers to only contain correctly received frames. In other words, we do not have access to a representative set of channel realizations for higher distances as the received packets with RSSI values below the carrier-sensitivity threshold had been truncated from the data points. Hence, to avoid the accuracy of model being affected, despite the existence of partial data points for the whole field test range, the model derivation is performed for distances up to 800 meters and 400 meters for touchstone and 400-car scenarios, respectively. It is noteworthy that precise propagation models for ranges up to around 1 km are useful and important to conduct proper and perfect simulation and analysis in different VANET research fields. However, based on provided results in one of our previous works, [10], which shows an acceptable level of model accuracy based on two special cases of $\alpha$-$\mu$ model, i.e. Nakagami-$m$ and Weibull distributions for ranges up to 1 km, it is reasonable to rely on $\alpha$-$\mu$ model for distances beyond 400 m, as well. In addition, it is worth mentioning that 400 meters is an acceptable range for the purpose of safety applications design and analysis, based on several technical documents issued by National Highway Traffic Safety Administration (NHTSA). For instance, NHTSA Technical Report on "Vehicle-to-Vehicle Communications: Readiness of V2V Technology for Application" [61] explicitly indicates 300 meters as the required transmitting range from a standard V2V communication technology such as DSRC to facilitate the identification of potential crashes by developing appropriate cooperative safety applications. The same fact has been reported in several other NHTSA technical documents such as [62], and [63].

### A. Touchstone Scenario

As highlighted in the previous section, the value of $\epsilon_r$ is determined from the touchstone scenario. Furthermore, the validity of our proposed approach is verified by comparing its results with regression based methods in this scenario as a benchmark. Based on the ideal channel condition assumption in single transmitter-receiver scenarios, path-loss follows a two-ray model with exponent value of $\gamma = 2$. The statistical mode of the distribution of obtained $\epsilon_r$ values for all distance pairs is shown in Figure 4 with respect to bin size of 0.01. Therefore, the value of $\epsilon_r = 1.01$ is chosen for our measurement campaign environment.

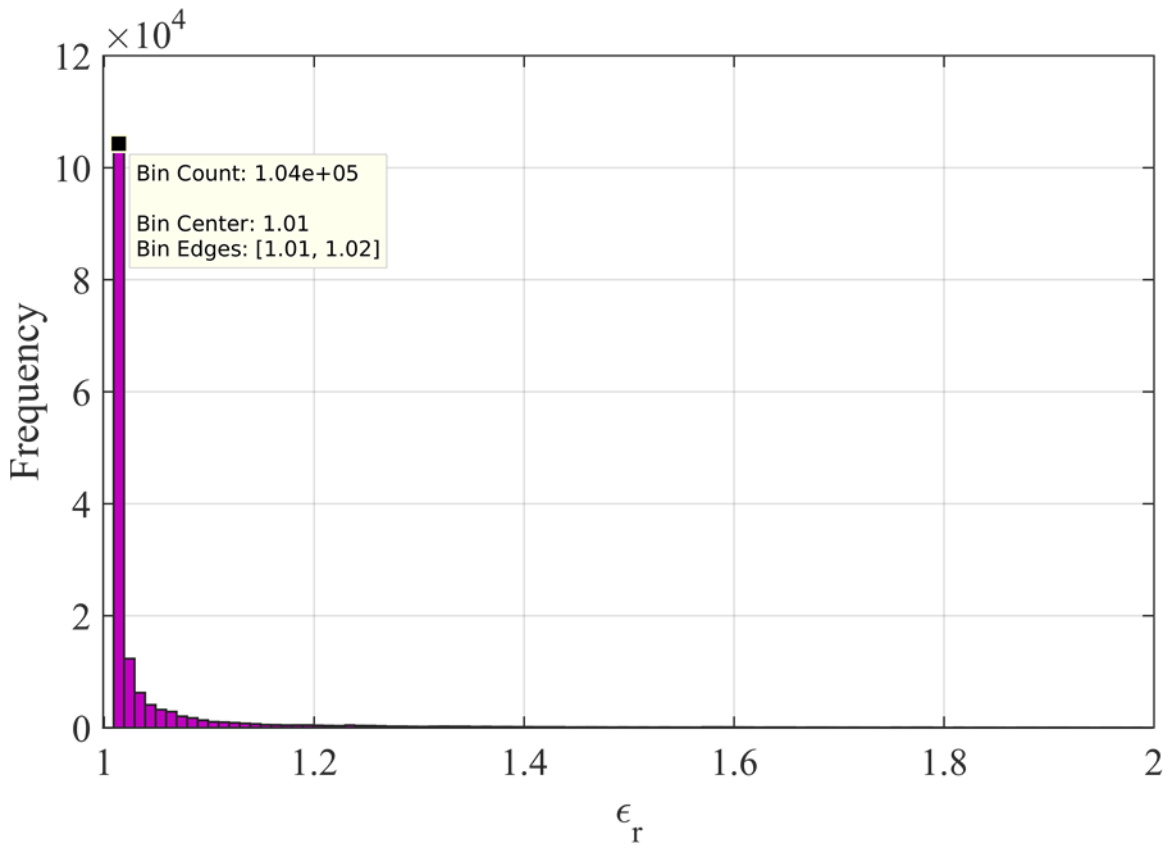

Figure 4- Histogram of calculated $\epsilon_r$ values from touchstone scenario

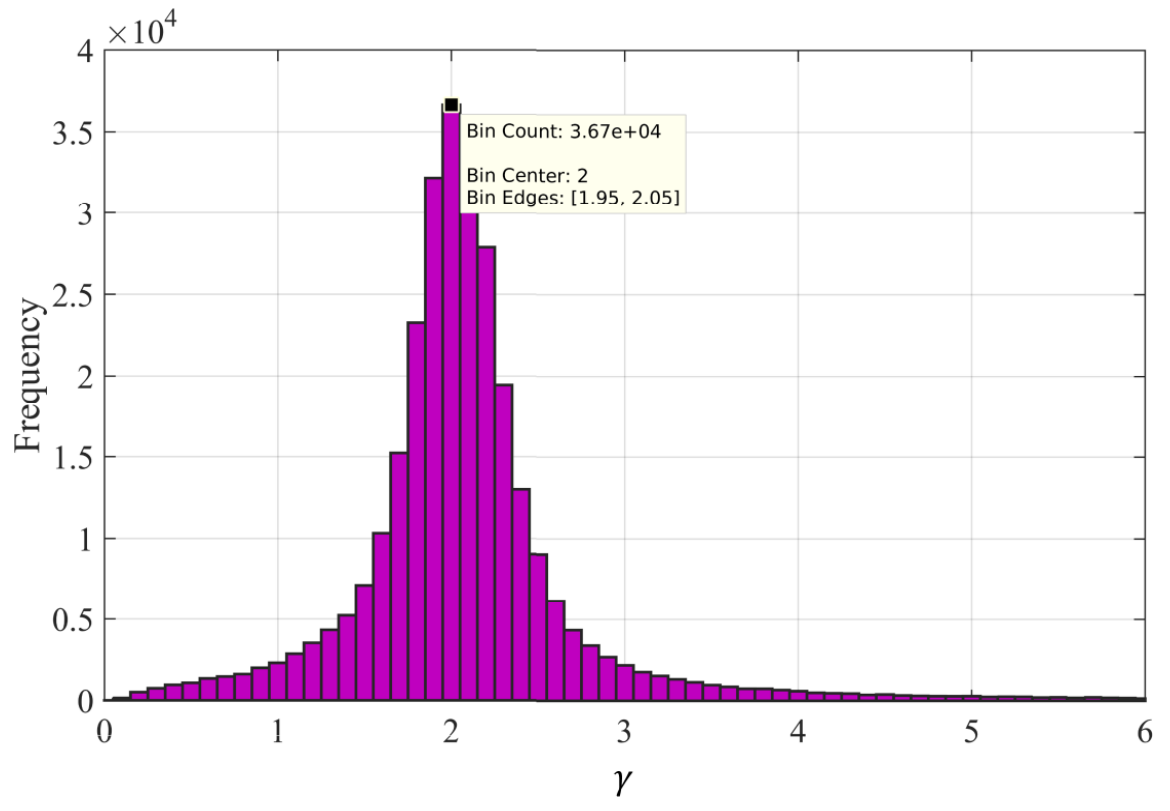

Figure 5- Histogram of calculated $\gamma$ values from touchstone scenario

The same method is used to verify the value of $\gamma = 2$ based on the figured value of $\epsilon_r = 1.01$. The statistical mode of the distribution of obtained values of $\gamma$ is shown in Figure 5. The value of $\gamma = 2$ is consistent with our assumption. It should be noted that this is not a trivial result since the regression method is a brute force approach, not a bijective mapping, which does not necessarily guarantee a one-to-one correspondence between $\gamma$ and $\epsilon_r$.

Moreover, the aforementioned regression method with minimum mean square error criteria is used to simultaneously calculate the values of $\gamma$ and $\epsilon_r$ in touchstone scenario to verify the validity of our approach. The values of $\epsilon_r = 1.01$ and $\gamma = 2.01$ are computed for regression method. Average RSSI data points of touchstone scenario and analytical two-ray path-loss models with parameters derived from two approaches are illustrated in Figure 6. Note that the RSSI values are averaged over one-meter bins.

### B. 400-car Scenario

#### 1) Path-loss Model Evaluation

Using the aforementioned method, the value of $\gamma$ is calculated for 400-car scenario based on the value of $\epsilon_r = 1.01$. Figure 7 shows the histogram of the distribution of derived values of $\gamma$ along with the most frequent bin, which is the chosen $\gamma = 2.1$.

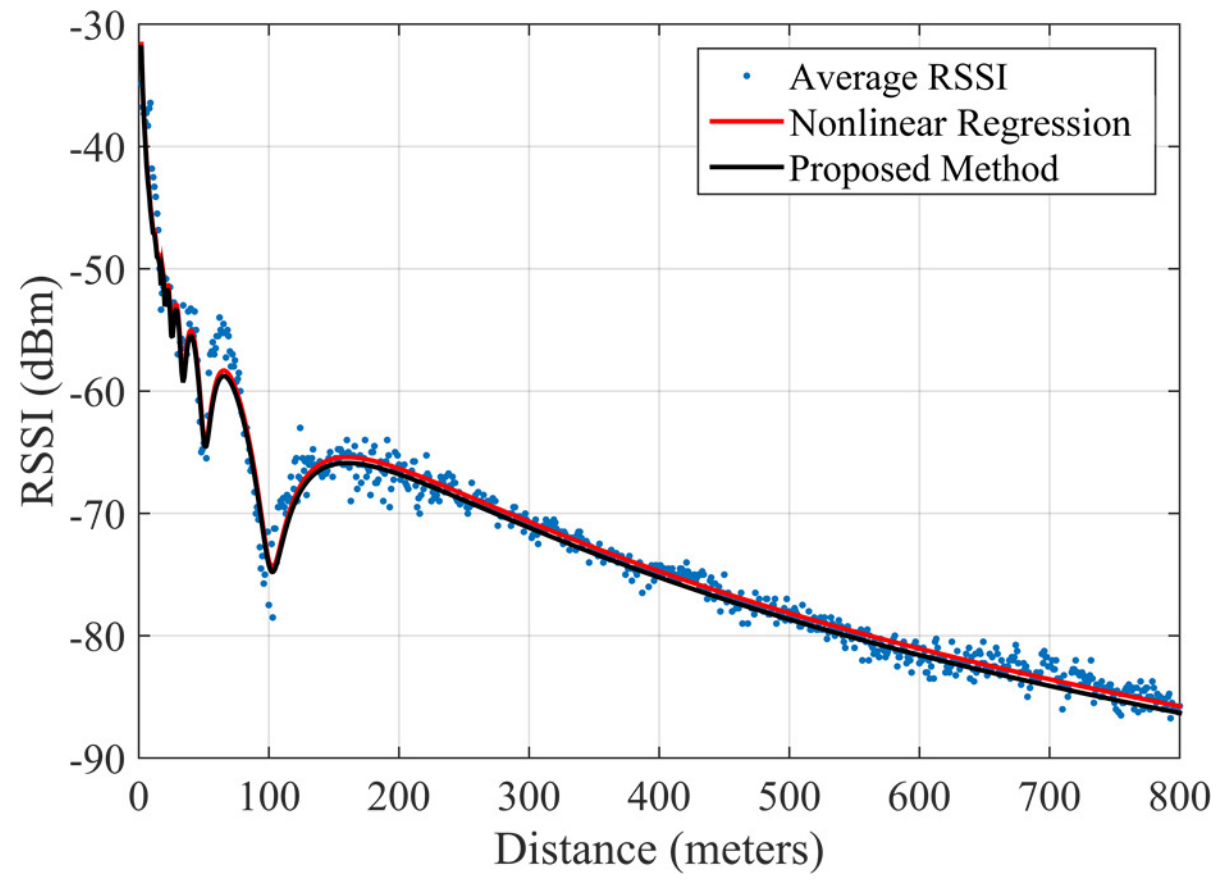

Figure 6- Validation of the proposed approach using one-meter bin averaged RSSI values of touchstone scenario

*2) Fading Model Evaluation*

The Path-loss model found in Step 1, with the values of $\epsilon_r = 1.01$ and $\gamma = 2.1$, is subtracted from the original RSSI data points to find the nondeterministic part of data. The aforementioned log-moment estimators are employed to derive the $\alpha$-$\mu$ fading model parameters. The range of 400 meters is divided into equal size distance bins for which a representative set of empirical data is available, and the best set of fading model parameters are estimated for each bin. To avoid overfitting of the estimated model parameters to our dataset, different binning values have been tried in this work. We derived $\alpha$-$\mu$ distribution parameters for 5-m, 10-m, 20-m, 40-m, 80-m, 100-m and 200-m binning sizes which are equivalent to 80, 40, 20, 10, 5, 4 and 2 sets of model parameters for the whole 400 meters range, respectively. For instance, these estimated parameter values for 10-m, 40-m, 100-m and 200-m binning sizes are depicted in Figure 8 and Figure 9.

In order to demonstrate the accuracy of the proposed model and evaluate its goodness-of-fit, the same number of data points for each distance bin are regenerated using an $\alpha$-$\mu$ distribution with the estimated parameters of that bin.

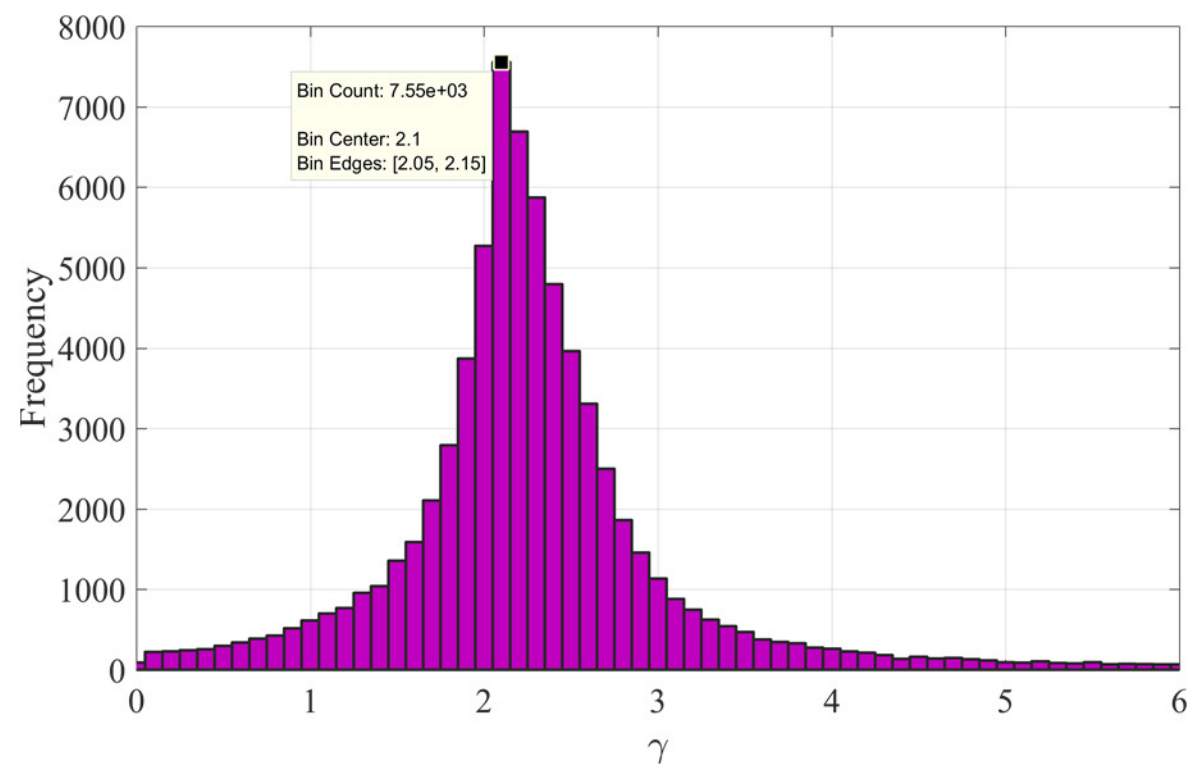

Figure 7- Histogram of calculated $\gamma$ values from 400-car scenario

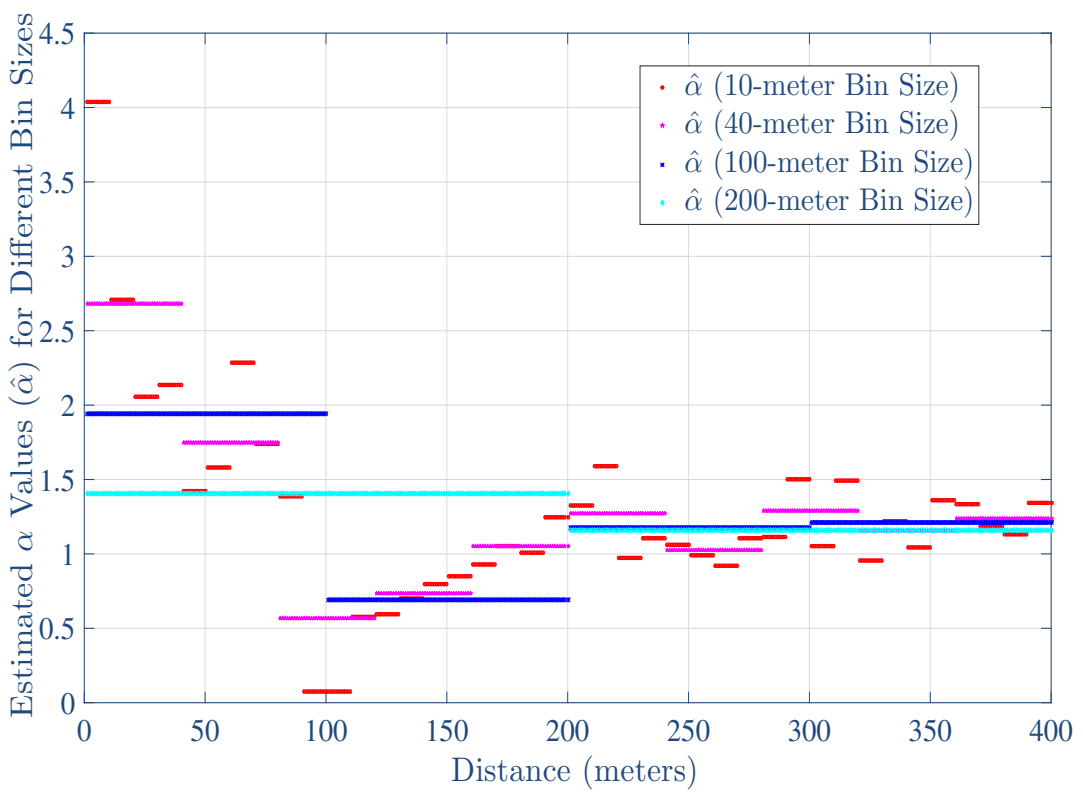

Figure 8- Estimated values for the first parameter of $\alpha$-$\mu$ fading distribution ($\hat{\alpha}$) for different binning sizes

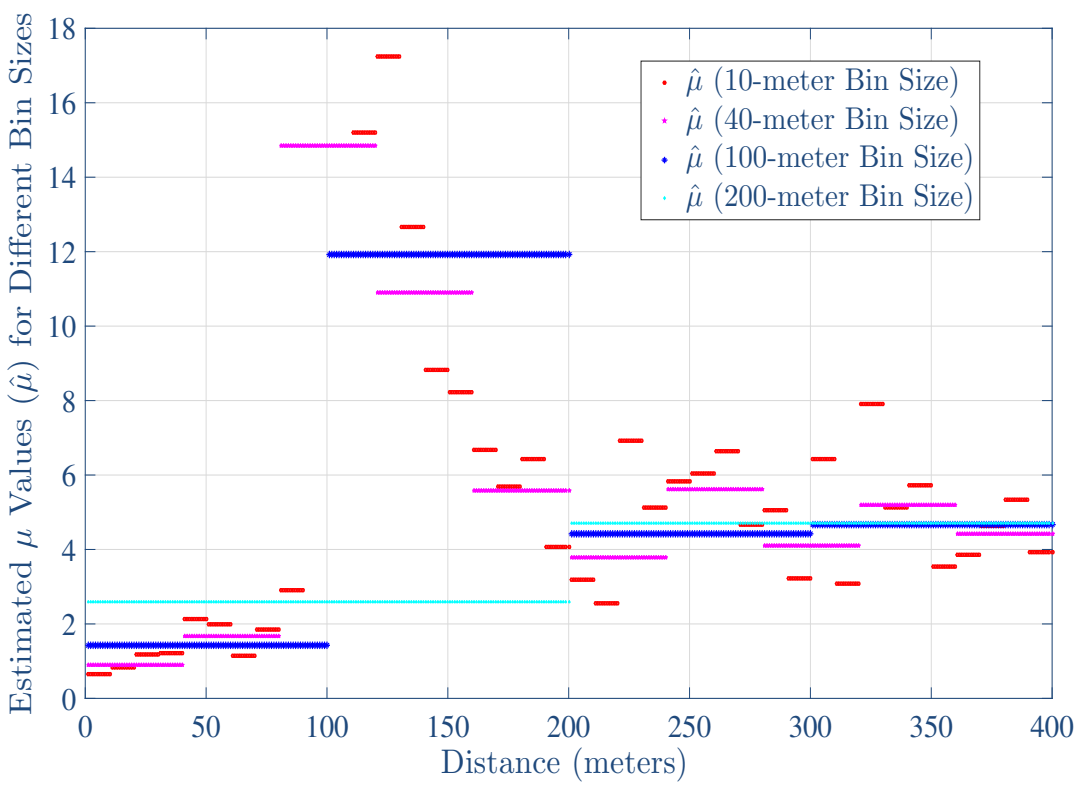

Figure 9- Estimated values for the second parameter of $\alpha$-$\mu$ fading distribution ($\hat{\mu}$) for different binning sizes

For instance, the regenerated RSSI values of each 5-m bin size along with empirical RSSI data points are represented as box plots in Figure 10. The field box plots are shown in blue and the regenerated points based on the derived model are depicted in red. The central black mark of each box is the median, the 25th and 75th percentiles are the edges of the box, and the whiskers extend to the most extreme non-outlier data points correspond to approximately 99.3 percent coverage if the data points are assumed to be normally distributed.

Kolmogorov-Smirnov test (K-S test) is a hypothesis testing approach for goodness-of-fit evaluation of a theoretic distribution with an empirical dataset [64]. K-S test is widely employed in the literature due to its two important characteristics [42], [53]. First, K-S test statistic is independent of the actual CDF of the empirical data. Second, binning size does not affect the validity of the K-S test results. The empirical CDF, $F_E(x)$, is defined as

$$F_E(x) = \frac{1}{n}\sum_{i=1}^{n} I(x_i \leq x), \tag{21}$$

where $I(x_i \leq x)$ stands for the indicator function which equals to one when the $i^{th}$ empirical data sample $(x_i)$ is less than or equal to $x$ and zero otherwise, and $n$ is the size of whole empirical dataset. In other words, $F_E(x)$ works as a normalized counter of empirical dataset elements based on the CDF

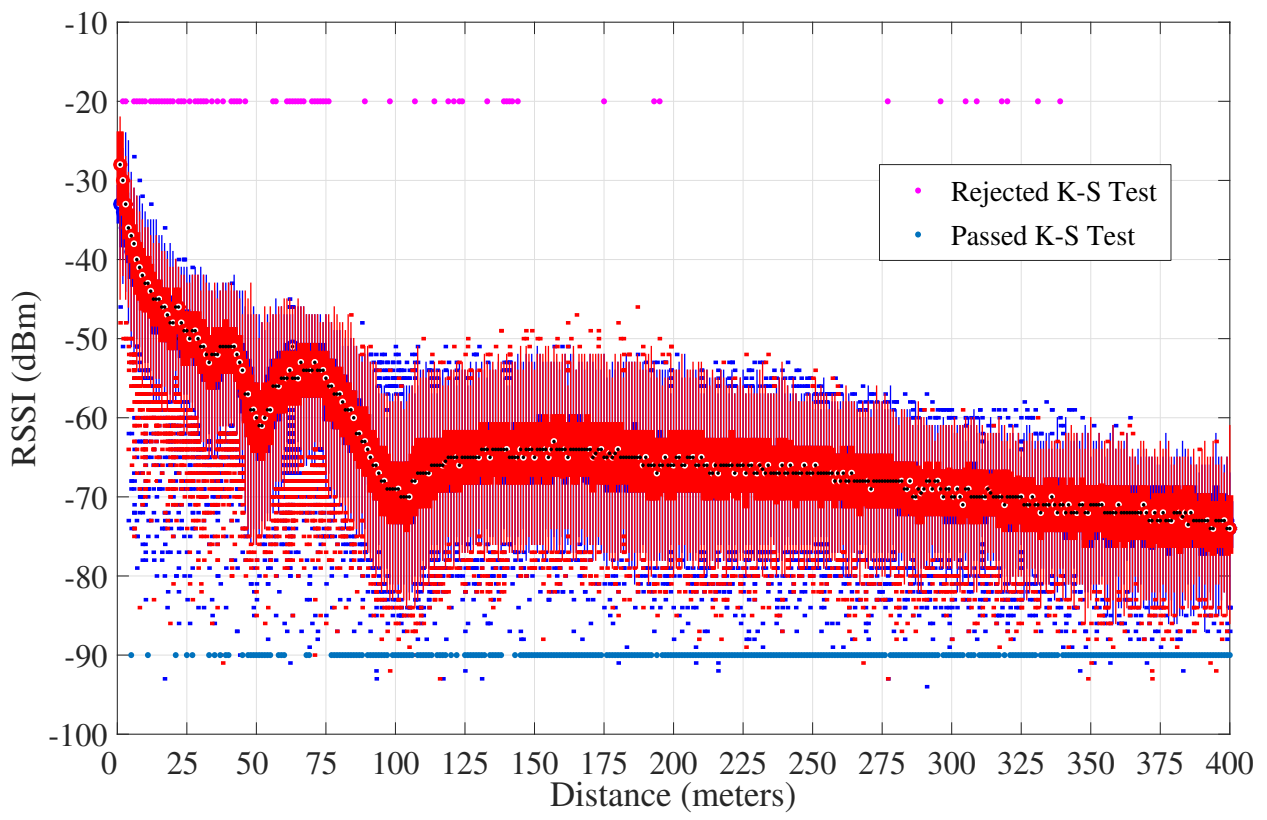

Figure 10- Box plot of empirical data (blue) versus regenerated points from $\alpha-\mu$ distribution (red) in 400-car scenario along with K-S test result (5-meter bin size)

definition. The main concept behind the K-S test is the similarity judgment of empirical CDF and hypothesized CDF based on the maximum value of their absolute distance or vertical discrepancy. The maximum vertical discrepancy between these two CDFs is known as the K-S test statistic and denoted as $D$:

$$D = \max_{x} |F(x) - F_E(x)|, \quad (22)$$

where $F(x)$ is the hypothesized CDF for the empirical data. K-S test results for the 400-car scenario RSSI values versus the $\alpha$-$\mu$ distribution are also illustrated in Figure 10 for 5-m binning size. Bin-wise pass and fail results of performed K-S test with the test significance level of 0.01 are derived and analyzed to verify our claims. These results for 5-m binning size are also depicted in Figure 10 in navy blue and purple, respectively.

At this point another statistical analysis layer is added on top of our K-S test results which enables us to derive concrete conclusions about dominant better performance of $\alpha$-$\mu$ distribution to model vehicular channel fading compared to Nakagami-$m$, as the most adopted distribution in this context. To this end, the same procedure which was conducted for $\alpha$-$\mu$ distribution, i.e. parameter estimation for all equal size bins using empirical RSSI values and then regenerating data points using these estimated parameters, is repeated for Nakagami-$m$ distribution as well. These calculations are performed for all of the mentioned binning sizes, i.e. 5-m, 10-m, 20-m, 40-m, 80-m, 100-m, and 200-m. As mentioned earlier, the number of regenerated data points for each bin is equal to its corresponding bin in empirical data to perform a valid K-S test. Well-known Maximum Likelihood (ML) estimation method is utilized in this work to estimate Nakagami-$m$ parameters. The estimated Nakagami-$m$ shape and scale parameters, $\hat{m}$ and $\hat{\omega}$, for 10-m, 40-m, 100-m, and 200-m binning sizes are represented in Figure 11 and Figure 12, respectively.

In order to make the comparison conclusions statistically meaningful, random point regeneration using the estimated parameters, is repeated 100 times for each distribution in each bin. This repetition is performed for all binning size cases. In the next step, 1-meter bin wise K-S tests are performed to compare each of these sets of randomly regenerated points with the field data. These K-S tests provide us with a sequence of

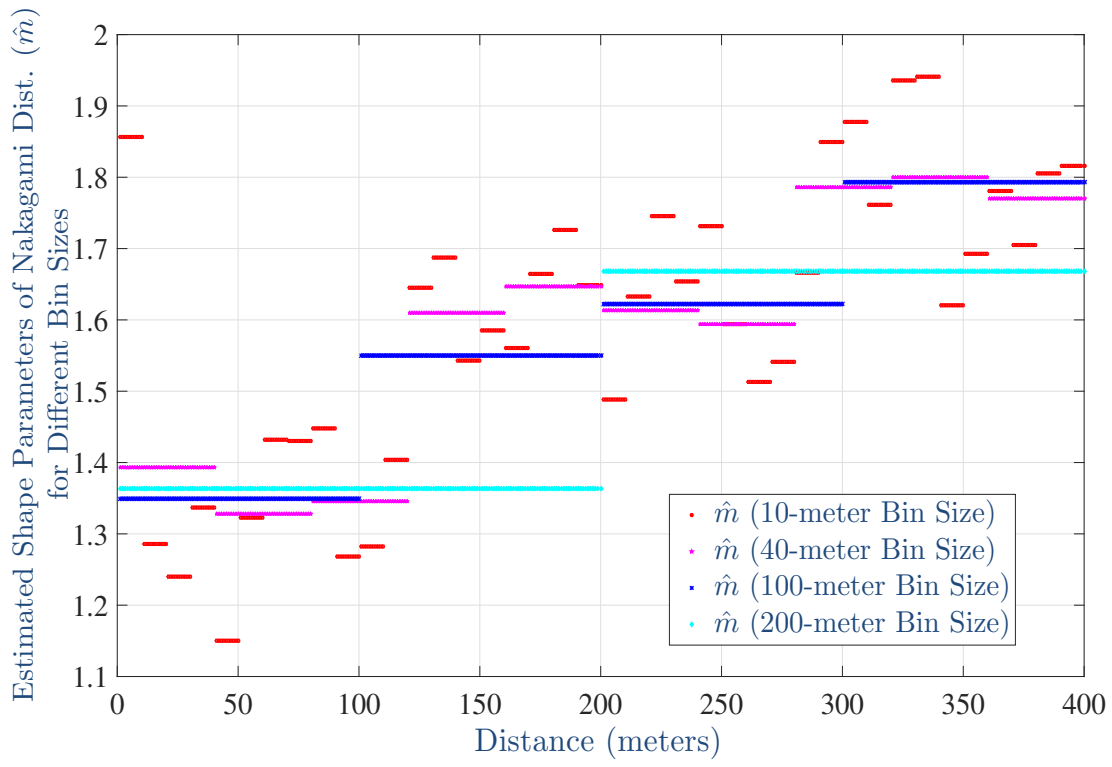

Figure 11- Estimated values for the shape parameter of Nakagami-$m$ fading distribution ($\hat{m}$) for different binning sizes

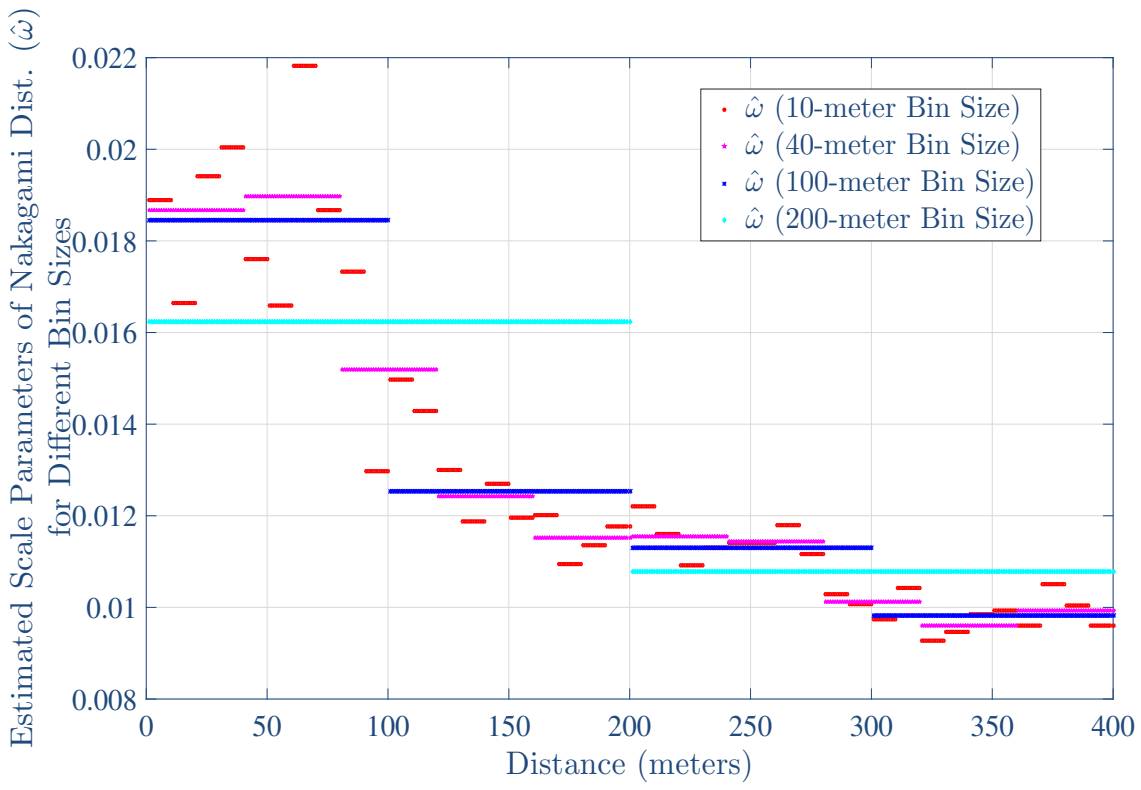

Figure 12- Estimated values for the scale parameter of Nakagami-$m$ fading distribution ($\hat{\omega}$) for different binning sizes

100 elements for each distribution in each binning size case. Elements of these sequences are the total number of 1-meter bins in which K-S test is passed under the circumstances, i.e. utilized distribution and binning size, which that sequence is generated. Thus, each element could range from 0 to 400. For instance, the achieved two sequences for $\alpha$-$\mu$ and Nakagami-$m$ distributions in 80-meter binning size case, is depicted in Figure 13. Moreover, the average values of all sequences from different examined binning sizes are shown in Figure 14. It could be obviously concluded from these figures that $\alpha$-$\mu$ distribution is always a better fit to the field dataset in comparison with Nakagami-$m$ in terms of successful K-S tests, independent of the chosen value for parameter estimation binning size. However, to explain this observational conclusion as a statistically meaningful statement, appropriate statistical tests are performed on each pair of derived 100-element sequences to demonstrate that they are significantly different. The results of these tests are presented in the appendix as Table I. First, Anderson-Darling (A-D) test is performed on each sequence to reveal its normality. For cases which A-D test cannot reject the null hypothesis, i.e. normality assumption, homoscedasticity test is also required to evaluate the variance equality assumption for two sequences. In situations that both sequences are recognized as normal with equal variances, unpaired T-test, which is a parametric test, is utilized while the non-parametric Wilcoxon test is performed for the remaining

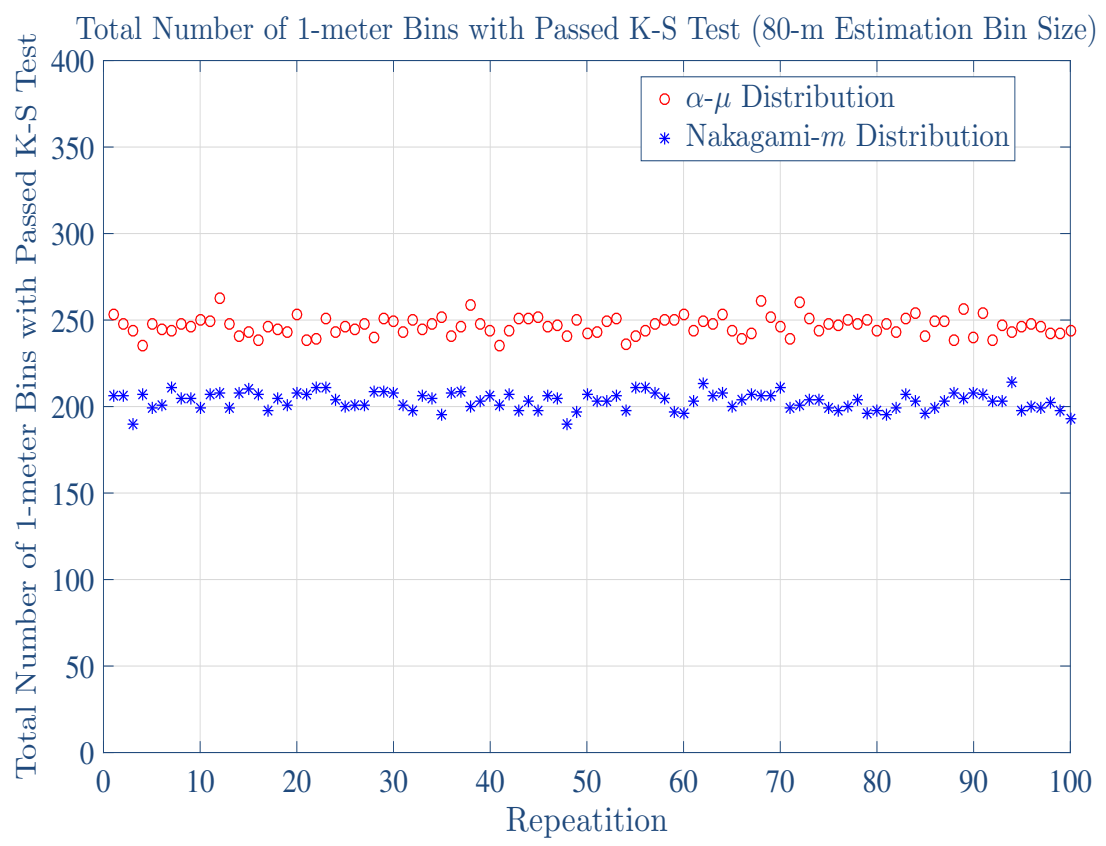

Figure 13- Total number of 1-meter bins, out of 400 bins, with successful K-S test result for $\alpha$-$\mu$ and Nakagami-$m$ regenerated points when compared to empirical data points (80-m parameter estimation binning size)

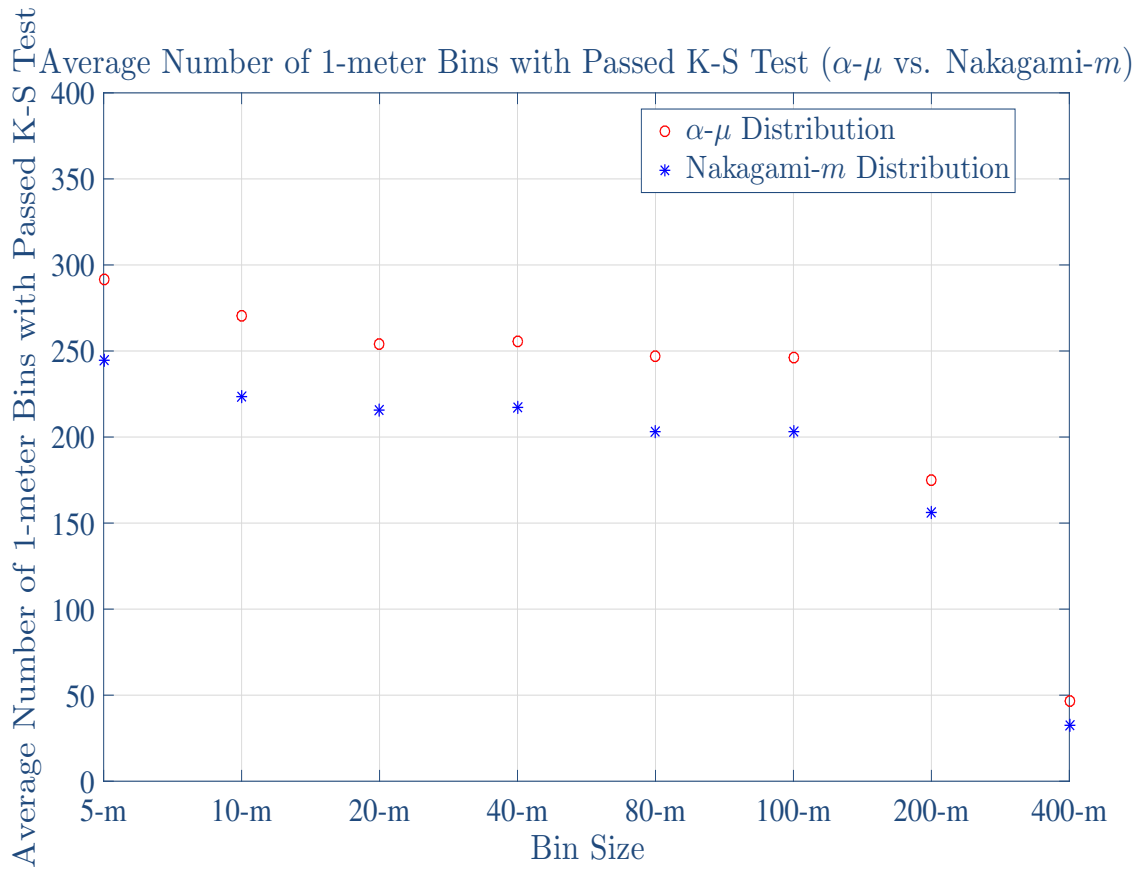

Figure 14- Average number of 1-meter bins, out of 400 bins, with successful K-S test result for $\alpha$-$\mu$ and Nakagami-$m$ regenerated points when compared to empirical data points (all parameter estimation binning size cases)

cases. All reported results in Table I are obtained based on 0.05 significance level. According to the reported h- and p-values in last two columns of Table I, similarity of two sequences are strongly rejected for all situations. These results firmly support our claim of better performance of $\alpha$-$\mu$ fading model in comparison with Nakagami-$m$, on dense vehicular networks.

We have also implemented both large and small-scale propagation models in ns-3, which is a realistic network simulator, to demonstrate the effectiveness of our proposed model. Figure 15 shows the transmission-reception procedure in ns-3.

The box plot comparison of simulation results versus field data is shown in Figure 16. This figure, which is generated based on 5-meter binning size for $\alpha$-$\mu$ parameter estimation, shows a good qualitative match between field data and the reconstructed data given by our model through its ns-3 implementation for the median curves, which is formed by central black mark of each box, and also for the 25$^{th}$ and 75$^{th}$ percentiles, the edges of the box. Moreover, whiskers, which are extended to the most extreme non-outlier data points correspond to approximately 99.3 percent coverage, have a good level of concordance, as well. However, we believe that a statistical comparison between the ns-3 generated data points and the field test data is not a meaningful and fair measure of the physical layer model accuracy since the output RSS samples of ns-3 simulations are outcome of various implemented models of the IEEE 802.11p protocol stack. For instance, the IEEE 802.11p MAC model and DSRC receiver model have significant impact on the received RSS samples. The effect of these models in ns-3 cannot be isolated from the effect of channel propagation model. However, as mentioned before, a qualitative comparison of the field test and the ns-3 results show that they are in good agreement.

## VI. Concluding Remarks

We investigated the modeling of large- and small-scale signal variations in vehicular networks using a 400-car test dataset. We first proposed an approach to remove the effect of fading on deterministic part of the large-scale model. The accuracy of our approach was verified using a touchstone scenario, which is a single transmitter-receiver scenario. Two-ray model was then utilized for path-loss characterization and its parameters were derived from the empirical data based on our proposed method. Furthermore, we studied the usage of $\alpha$-$\mu$ distribution to model the fading behavior of vehicular networks for the first time, and verified its veracity by the K-S goodness-of-fit test. A large RSSI dataset from a measurement campaign was used to evaluate our claims. Moreover, ns-3 was used to show the outcome of the proposed model in the presence of upper network layers.

Based on the presented results of this paper, utilizing more general distributions, such as $\alpha$-$\mu$ distribution, is observed to be a promising approach to model the fading behavior of vehicular environments and gives considerable better results compared to the mostly adopted ones such as Nakagami$-m$. Therefore, we are planning to extend the proposed approach for other dense vehicular environments such as intersections and tunnels.

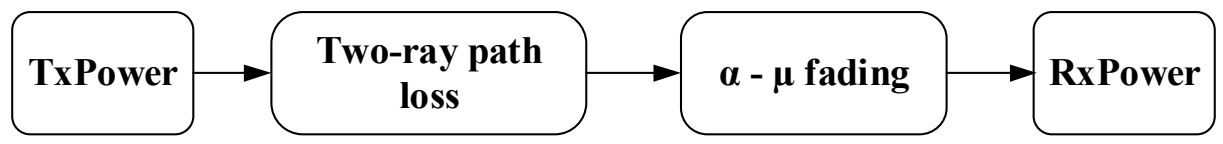

Figure 15- Steps of model implementation in ns-3

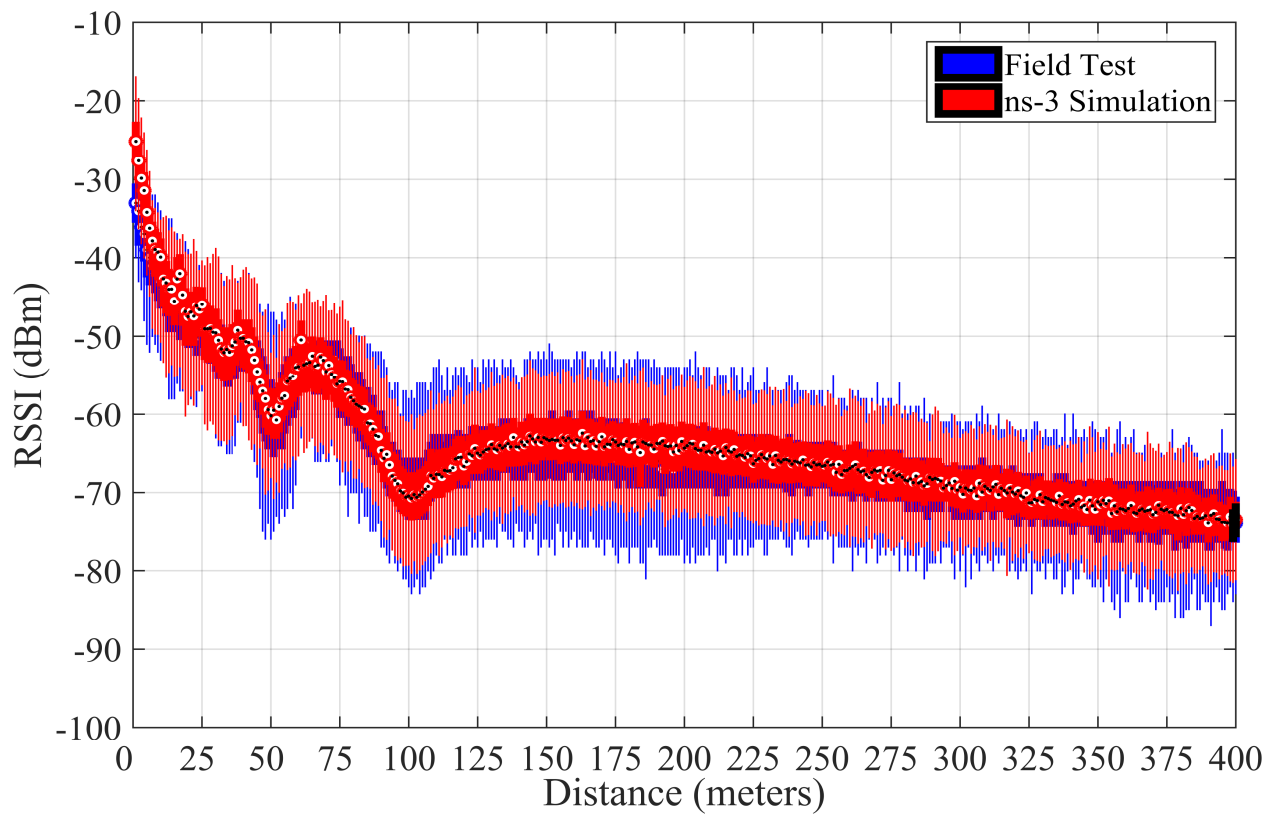

Figure 16- Box plot of empirical data (blue) versus ns-3 simulation results (red) in 400-car scenario.

## APPENDIX

TABLE I
STATISTICAL COMPARISON OF K-S TEST RESULTS FOR $\alpha$-$\mu$ AND NAKAGAMI-$m$ DISTRIBUTIONS VERSUS FIELD DATA POINTS, BASED ON REPETITIONS IN RANDOM POINTS REGENERATION

| Test / Bin Size | | Anderson-Darling Test | | Homoscedasticity Test | Unpaired Two-Sample T-Test | Wilcoxon Rank Sum Test |
|---|---|---|---|---|---|---|
| | | Nakagami-$m$ | $\alpha$-$\mu$ | | | |
| 5-meter | p-value | 0.1838 | 0.0917 | 0.3809 | 1.3496e-140 | - |
| | h-value | 0 | 0 | 0 | 1 | |
| 10-meter | p-value | 0.2427 | 0.1133 | 0.1570 | 1.5409e-141 | - |
| | h-value | 0 | 0 | 0 | 1 | |
| 20-meter | p-value | 0.0343 | 0.3304 | - | - | 2.3174e-34 |
| | h-value | 1 | 0 | | | 1 |
| 40-meter | p-value | 0.8708 | 0.0843 | 0.0866 | 1.1144e-111 | - |
| | h-value | 0 | 0 | 0 | 1 | |
| 80-meter | p-value | 0.0314 | 0.1989 | - | - | 2.3183e-34 |
| | h-value | 1 | 0 | | | 1 |
| 100-meter | p-value | 0.3201 | 0.2814 | 0.0874 | 7.7113e-135 | - |
| | h-value | 0 | 0 | 0 | 1 | |
| 200-meter | p-value | 0.0265 | 0.0950 | - | - | 4.1759e-34 |
| | h-value | 1 | 0 | | | 1 |
| 400-meter | p-value | 0.0036 | 0.0198 | - | - | 2.1514e-34 |
| | h-value | 1 | 1 | | | 1 |

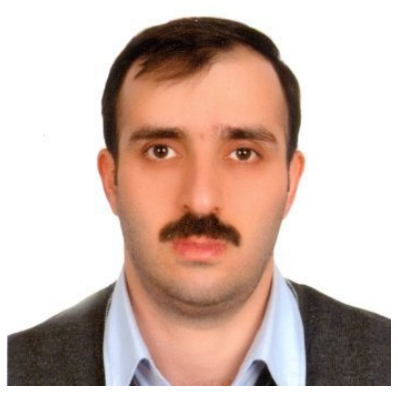

**Hossein Nourkhiz Mahjoub** is a Ph.D. student in electrical engineering with the Department of Electrical and Computer Engineering, University of Central Florida, Orlando, FL, USA. He received his B.S. and M.S. degrees in electrical engineering (Systems Communications) from University of Tehran, Tehran, Iran, in 2003 and 2008, respectively. He has more than nine years of work experience in the telecommunications industry before starting his Ph.D. in 2015. His research interests include wireless channel modeling, stochastic systems analysis, and vehicular ad -hoc networks.

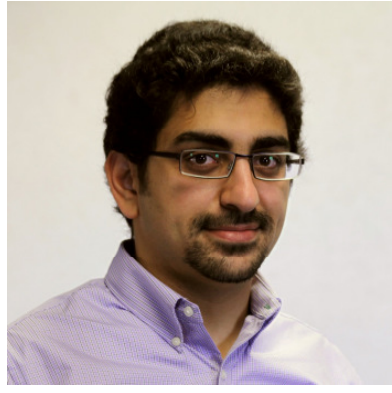

**Amin Tahmasbi-Sarvestani** received the B.S. degree in software engineering from the University of Isfahan, Isfahan, Iran, in 2009 and the M.S. degree in computer engineering (artificial intelligence) from Sharif University of Technology, Tehran, Iran, in 2011. He is currently working toward the Ph.D. degree in computer science with the Lane Department of Computer Science and Electrical Engineering, West Virginia University, Morgantown, WV, USA.
His research interests include transportation cyber-physical systems and vehicular networks.

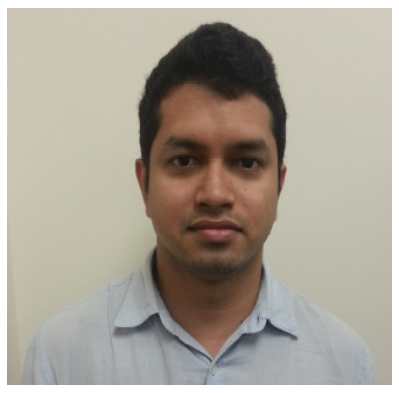

**S M Osman Gani** is currently pursuing PhD degree in Computer Engineering at University of Central Florida. He received his B.S. in Computer Science and Engineering from Bangladesh University of Engineering and Technology and M.S. in Computer Science from West Virginia University. His current research interests include wireless communications in vehicular networks and intelligent transportation systems.

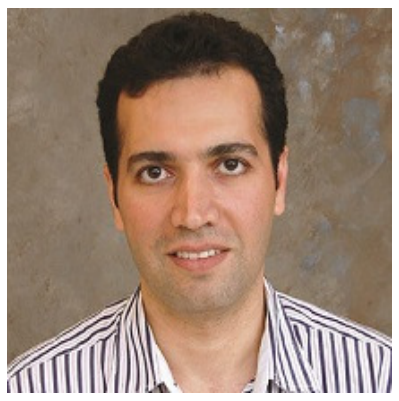

**Yaser P. Fallah** received his PhD from the University of British Columbia, Canada, in 2007. He is currently an associate professor of Electrical and Computer Engineering at University of Central Florida. From 2011 to 2016, he was an assistant professor at West Virginia University. Prior to joining WVU, he worked as a research scientist at the University of California Berkeley, Institute of Transportation Studies (2008 -2011). Dr. Fallah is an editor of IEEE Trans. On Vehicular Technology; he has served as the program chair of IEEE Wireless Vehicular Communication symposia in 2011 and 2014 and demo/poster co -chair of IEEE Vehicular Networking Conference in 2016. His current research, sponsored by industry and USDoT projects as well as NSF CAREER award, focuses on intelligent transportation systems and involves analysis and design of automated and networked vehicle safety systems.